\documentclass[12pt]{article}
\usepackage[dvips]{graphicx}

\usepackage{epsfig,amsmath,amssymb,verbatim,mathrsfs}

\def\beq{\begin{eqnarray}}
\def\eeq{\end{eqnarray}}
\def\bea{\begin{eqnarray}}
\def\eea{\end{eqnarray}}
\def\blfootnote{\xdef\@thefnmark{}\@footnotetext}

\newcommand{\D}{\text{\tiny D}}

\newcommand{\ri}{\text{\tiny R}}
\newcommand{\li}{\text{\tiny L}}
\newcommand{\nl}{\text{\tiny N}}

\newcommand{\be}{\begin{equation}}
\newcommand{\ee}{\end{equation}}
\newcommand{\nubb}{0\nu\beta\beta}

\begin{document}
\begin{titlepage}
\noindent

\setlength{\baselineskip}{0.2in}
\flushright{NIKHEF/2011-013\\May 2011}
\vspace{0.4cm}

\begin{center}
  \begin{Large}
    \begin{bf}
 Neutrino Mass and $\boldsymbol{\mu\rightarrow e+ \gamma}$ from a Mini-Seesaw
\end{bf}
  \end{Large}
\end{center}
\vspace{0.2cm}

\begin{center}
\begin{large}
{
Michael Duerr,$^{(a),}$\footnote{Email: michael.duerr@mpi-hd.mpg.de}
  Damien~P.~George$^{(b),}$\footnote{Email: dpgeorge@nikhef.nl} and Kristian~L.~McDonald$^{(a),}$\footnote{Email: kristian.mcdonald@mpi-hd.mpg.de}}\\
\end{large}
\vspace{1cm}
  \begin{it}
$(a)$ Max-Planck-Institut f\"ur Kernphysik,\\
 Postfach 10 39 80, 69029 Heidelberg, Germany\vspace{0.5cm}\\
$(b)$ Nikhef Theory Group, Science Park 105,\\
1098 XG Amsterdam, The Netherlands\\\vspace{0.5cm}

\end{it}
\vspace{1cm}

\end{center}


\begin{abstract}
The recently proposed ``mini-seesaw mechanism'' combines naturally
suppressed Dirac and Majorana masses to achieve 
light Standard 
Model neutrinos via a low-scale seesaw. 
A key
feature of this approach
is the presence of multiple light (order GeV) sterile-neutrinos that mix with the
Standard Model. In this work we study the
bounds on these light sterile-neutrinos from processes like
$\mu\rightarrow e+\gamma$, invisible $Z$-decays, and neutrinoless double beta-decay. We
show that viable parameter space exists and that, interestingly,
key observables can lie just below current experimental
sensitivities. In particular, a motivated region of parameter
space predicts a $\mu\rightarrow e+\gamma$ branching fraction
 within the range to be probed by MEG.

\end{abstract}

\vspace{2cm}

\end{titlepage}

\setcounter{page}{1}

\vfill\eject

\section{Introduction}
The discovery of neutrino mass has provided a wealth of information regarding the flavour structure of Nature (for reviews see~\cite{Nu_review}; for the latest global fit see~\cite{Schwetz:2011qt}). 
However, key questions regarding the neutrino sector remain unanswered
and, in particular,  the underlying mechanism responsible for neutrino
 mass remains elusive. The
 seesaw mechanism provides a simple explanation for
the lightness of the known neutrinos~\cite{seesaw}. In its standard implementation one assumes
$M_R\gg m_{\D}\sim m_W$, where $m_{\D}$ 
 is the neutrino Dirac mass, $M_R$ is the singlet Majorana mass and
 $m_W$ is the $W$ boson mass. The result is a light Standard Model (SM) neutrino with mass $m_\nu\sim
 m_{\D}^2/M_R$ and a heavy sterile neutrino with mass $\sim
 M_R$.  Despite the simplicity of this approach, it offers little hope of being experimentally verified due to the invocation of the typically (very) large scale $M_R$.

A number of alternative mechanisms invoke new physics to explain
neutrino masses without relying on a large UV scale. These offer the
advantage of being experimentally verifiable and one may hope that the
requisite new physics appears at the TeV scale. This can happen in
models with radiatively-generated neutrino mass~\cite{Zee:1985id}, and
the mechanism of mass generation can even be connected to the weak
scale in models with Coleman-Weinberg symmetry
breaking~\cite{Meissner:2006zh}. Of course, Nature is likely not
concerned with our ability to experimentally ascertain its inner
workings.  Nonetheless it is interesting to explore alternative
approaches to neutrino mass that may lie within experimental reach;
particularly if they enable a natural realization and/or provide
insight into the observed flavour structures.

Beyond being SM singlets, right-handed neutrinos can form part of a
more complicated hidden sector, and may provide a link to the
dark/hidden sector of the Universe. This feature is used to advantage
in models that invoke non-standard properties of the gauge-singlet
neutrinos in an effort to understand the origin of neutrino mass. For
example,
interesting exceptions to the standard seesaw picture arise if
the right-handed neutrinos are bulk fields in an extra dimension~\cite{ArkaniHamed:1998vp} or if they are composite objects within  a strongly
coupled hidden sector~\cite{ArkaniHamed:1998pf}. 

A recent work has proposed a ``mini-seesaw mechanism'' in which naturally
suppressed Dirac and (sterile) Majorana masses are combined to achieve 
light SM neutrinos via a low-scale seesaw~\cite{McDonald:2010jm}. This
approach borrows elements from both the extra-dimensional scenario and
the composite right-handed neutrino approach: the sterile neutrinos
are bulk fields in a truncated slice of $AdS_5$ but, via the AdS/CFT
correspondence~\cite{Maldacena:1997re, ArkaniHamed:2000ds}, the model
has a dual 4D description in which the right-handed neutrinos are the
lightest composites of a strongly coupled hidden sector. A key feature
of this approach is the presence of a tower of light
(order GeV) sterile neutrinos that can be thought of as either Kaluza-Klein
modes or higher fermionic resonances of the CFT (with mass gap). In
this work we consider the detailed bounds on these light
sterile-neutrinos. We show that viable parameter space exists in which
SM neutrino masses can be successfully generated without recourse
to large (supra-TeV) scales. Furthermore, we show that it may be
possible to observe the effects of the sterile neutrinos
in forthcoming experiments: the most
stringent constraints come from the FCNC process
$\mu\rightarrow e+\gamma$ and, as we will show, the interesting region
of parameter
space corresponds to $BR(\mu\rightarrow
e+\gamma)\sim10^{-13}-10^{-12}$.
 This is right below the current experimental bound and within the
 region to be probed by the MEG experiment~\cite{MEG}. The MEG
 collaboration recently reported that the
 best value for the number 
 of signal events in their maximal likelihood
 fit is three~\cite{MEG}. Interestingly, this corresponds to $BR(\mu\rightarrow
e+\gamma)=3\times10^{-12}$ which, as we show, is obtained via
the mini-seesaw with neutrino Yukawa couplings approaching the
$\mathcal{O}(0.1)$ level.

Before proceeding we note that bulk gauge-singlet neutrinos
in Randall-Sundrum (RS) models~\cite{Randall:1999ee} were considered in~\cite{Grossman:1999ra} and bulk SM
fermions in~\cite{Gherghetta:2000qt}. Subsequent studies of neutrino
mass appeared in~\cite{Huber:2002gp} and for
an incomplete list of recent works in this active field
see~\cite{Gherghetta:2003he,Gripaios:2006dc}. Related work on
right-handed neutrinos within a strongly coupled CFT was undertaken in~\cite{vonGersdorff:2008is}. Our 
implementation within a sub-TeV scale effective theory
differs from these previous works. Other works have also considered a
low-scale warped hidden
sector~\cite{Flacke:2006re,McDonald:2010fe} and for another use of bulk sterile
neutrinos see~\cite{Pas:2005rb}. Light sterile-neutrinos
have  been studied extensively within the context of the
$\nu$MSM~\cite{Asaka:2005an,Asaka:2011pb} and for a discussion of the
low-scale (GeV) seesaw see~\cite{de
  Gouvea:2007uz}. Ref.~\cite{Kusenko:2009up} contains a
review on light steriles and a detailed analysis of the bounds on KK
neutrinos in ADD models appeared in~\cite{Ioannisian:1999cw}. A recent work has studied a
viable realization of the MEG signal in a SUSY
model~\cite{Fukuyama:2011rc} and other related recent works include~\cite{Blanchet:2009kk}. 

The layout of this paper is as follows. Section~\ref{sec:setup}
describes the model and outlines the basic neutrino spectrum. In
Section~\ref{sec:mass_eigenstates} the sterile neutrino spectrum
is considered in more detail and the mixing with the SM is
determined. The analysis of the bounds follows. We study the
neutrinoless double beta-decay rate and invisible $Z$-decays in
Sections~\ref{sec:nubb} and~\ref{sec:hid_z_decay} respectively, a
number of other bounds on active-sterile neutrino mixing in Section~\ref{sec:other_bounds} (e.g. collider
constraints), and the lepton number violating decay
$\mu\rightarrow e+\gamma$ in
Section~\ref{sec:mu_to_e_gamma}. Finally we discuss the prospects for
observing invisible Higgs decays in Section~\ref{sec:higgs_decay} and
conclude in Section~\ref{sec:conc}. 
\section{Light Neutrinos from a Mini-Seesaw\label{sec:setup}}
We consider a truncated RS model with a warped extra dimension described by the
coordinate $z\in[k^{-1},R]$. A UV brane of characteristic
energy scale $ k$ is located at
$z=k^{-1}$ and an IR brane with characteristic scale
$R^{-1}$ is located at $z=R$. The metric is given by
\beq
ds^2 = \frac{1}{(kz)^2}(\eta_{\mu\nu}dx^{\mu}dx^{\nu} -
dz^2)= G_{MN} dx^{M}dx^{N},
\label{bulkmetric}
\eeq
where $M,N,..$ ($\mu,\nu,..$) are the 5D (4D) Lorentz indices and
$k$ is the $AdS_5$ curvature. The characteristic IR scale is suppressed
relative to the curvature, 
as is readily seen using the 
proper coordinate for the extra dimension.\footnote{This is defined by $y= k^{-1}\log(kz)$, where $y\in[0,L]$ and $L=k^{-1}\log(kR)$, in terms of which the IR scale is exponentially suppressed, $R^{-1}=e^{-kL}k \ll k$.}
When sourced by a bulk cosmological constant and appropriate brane 
tensions the metric of Eq.~\eqref{bulkmetric}
is a solution to the 5D Einstein
equations~\cite{Randall:1999ee}. Furthermore, the length of the space is readily
stabilized~\cite{Goldberger:1999uk}.

\begin{figure}[ttt]
\begin{center}
        \includegraphics[width = 0.3\textwidth]{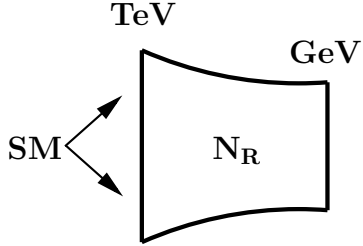}
\end{center}
\caption{Sketch of the ``Little Warped Space". The sterile neutrino $N_R$ propagates in a hidden warped space with order
  GeV IR scale and order TeV UV scale. Standard Model fields
  reside on the UV brane and have a suppressed coupling to the chiral zero
  mode neutrino, which is localized toward the IR brane.}
\label{fig:nu_lit_throat}
\end{figure}

We take the SM to be localized on
the UV brane, where the natural mass 
scale is $\sim k$, and accordingly take
$k\lesssim M_*\sim$~TeV, where $M_*$ is the cutoff scale. In addition, we consider three singlet fermions
propagating in the bulk. We label these by $N_R$ as the
zero modes will be right-chiral fields that we identify as gauge-singlet
neutrinos. The IR scale is nominally taken to
be
$R^{-1}\sim$~GeV. A sketch of the setup is given in
Figure~\ref{fig:nu_lit_throat}.  In analogy with the Little RS
model~\cite{Davoudiasl:2008hx} we can refer to this as a ``Little
Warped Space" (LWS). Such a truncated spacetime may seem
somewhat unusual at 
first sight. However, the setup can be thought of as an 
effective theory that describes the sub-TeV scale physics of a more
complete theory, enabling one to consider the effects of a light
warped/composite hidden-sector without having to specify the supra-TeV
physics. In general the supra-TeV effects will be encoded in
   UV localized effective operators. As with any sub-TeV scale
   effective theory, effective operators that break
   approximate/exact symmetries of the SM must be
   adequately suppressed  to
   ensure that problematic effects like rapid p-decay and excessive flavour
   violation do not occur.

The action for a bulk fermion $N_R$ in the
background (\ref{bulkmetric})  is\footnote{We have dropped the spin-connection terms which
cancel in $S_{\nl}$.}
\begin{eqnarray}
S_{\nl}&=&\int
d^5x\sqrt{G}\left\{\frac{i}{2}\bar{N}_R\Gamma^{B}e^A_{B}\partial_A
  N_R-\frac{i}{2}(\partial_A\bar{N}_R)\Gamma^{B}e^A_{B}N_R+ck\bar{N}_RN_R\right\},
\end{eqnarray}
where $\Gamma^{\mu,5}=\{\gamma^\mu,i\gamma^5\}$ are the 5D Dirac-gamma
matrices, $e^A_{B}$
is the f\"unfbein and we write the Dirac mass in units of the
curvature $k$. A Kaluza-Klein (KK) decomposition may be
performed as
\bea
N_R(x,z)=(kz)^2\sum_n \left\{\nu_L^{(n)}(x)f_L^{(n)}(z)+\nu_R^{(n)}(x)f_R^{(n)}(z)\right\},
\eea
and the bulk wavefunctions $f^{(n)}_{L,R}$ for the chiral components
of $N_R$ are readily found~\cite{Grossman:1999ra}:
\begin{eqnarray}
f_{L,R}^{(n)}(z)&=&\mp\frac{\sqrt{kz}}{N_{n}}\left\{J_{\alpha_{\li,\ri}}(M_{n}z)+\beta_{n}Y_{\alpha_{\li,\ri}}(M_{n}z)\right\},
\end{eqnarray} 
where the order of the Bessel functions is
$\alpha_{\li,\ri}=|c\pm1/2|$. The boundary
conditions force one chirality to be odd and without loss of
generality we take this
to be the left-chiral field. The single massless mode in the
spectrum then has right-chirality. 

 The KK-expanded Lagrangian is
\begin{eqnarray}
S_{\nl}&=&\sum_{n}\int
d^4x\left\{i\bar{\nu}^{(n)}\gamma^{\mu}\partial_\mu
  \nu^{(n)}-M_n\bar{\nu}^{(n)}\nu^{(n)}\right\},
\end{eqnarray}
where $\nu^{(n)}=\nu_L^{(n)}+\nu_R^{(n)}$ is a Dirac fermion with KK
mass $M_n$  for
$n>0$ and $\nu^{(0)}=\nu^{(0)}_R$ is the massless right-chiral zero
mode, whose bulk profile is
\bea
f^{(0)}_R(z)=\sqrt{\frac{k(1+2c)}{(kR)^{1+2c}-1}}\left(kz\right)^{c}.
\eea
The dimensionless mass parameter $c$ is seen to control the localization
of the zero mode. We will be interested in IR localization
with $c\simeq1$, in order to reduce the wavefunction overlap of the zero
mode with UV localized SM neutrinos and suppress
the Dirac mass below the weak scale.

The KK masses are determined by enforcing the boundary conditions (Dirichlet/Neumann for $f_{L/R}^{(n)}$), and must satisfy
 $J_{\alpha_{\li}}(M_n R)\simeq0$, giving $M_n\simeq(n+c/2)\pi R^{-1}$ for
$n>0$. For $c\simeq 1$ these masses are $M_nR\simeq 4.6,\ 7.7,\ 10.8\ \dots$
The boundary values of the wavefunctions for the right-chiral modes will be important in the following;
for $c\simeq1$ and $M_n<k$ these are
\bea
f_{R}^{(n)}(k^{-1})&\simeq&\frac{1}{\Gamma(c+1/2)}\sqrt{\frac{2\pi}{R}}\left(\frac{M_n}{2k}\right)^c\
,\nonumber\\
f_{R}^{(n)}(R)&\simeq&(-1)^n\sqrt{\frac{2}{R}}\ .
\eea

At this stage the spectrum
consists of a single Weyl neutrino and a tower of Dirac neutrinos with
masses $M_n\sim n\pi/R$. We now introduce
lepton number violation in the form of a marginal
operator on the IR brane:
\begin{eqnarray}
S_{\nl}&\rightarrow&S_{\nl} +\frac{\lambda_{\nl}}{2}\int
d^5x\sqrt{-g_{ir}}\left\{\bar{N}_R^cN_R
  +\mathrm{H.c.}\right\}\delta(z-R)\nonumber\\
&=&\sum_{m,n}\int
d^4x\left\{i\bar{\nu}^{(n)}\gamma^{\mu}\partial_\mu
  \nu^{(n)}-M_n\bar{\nu}^{(n)}\nu^{(n)}+\frac{M_{mn}}{2}(\bar{\nu}_R^{(m)})^c\nu^{(n)}_R+\mathrm{H.c.} \right\},
\eea  
where the effective Majorana masses are
\bea
M_{mn}&=& \left. \lambda_{\nl}f_R^{(m)}f_R^{(n)} \right|_{z=R}.
\eea
Note that $\nu_L^{(n)}$ does not acquire a boundary Majorana mass as $\left. f_L^{(n)} \right|_R=0$. For IR localization the zero mode Majorana mass takes a particularly simple form:
\bea
M_{00}&\simeq& (1+2c)\frac{\lambda_{\nl}}{R},\label{maj_zero_mode}
\eea
and the Majorana masses for the $m,n>0$
modes can be approximately related to that of the zero mode:
\bea
|M_{0n}|\simeq \sqrt{\frac{2}{2c+1}}M_{00}\quad\mathrm{and}\quad |M_{mn}|\simeq
\frac{2M_{00}}{(2c+1)}\quad\mathrm{for}\quad m,n>0\ .\label{maj_higher_mode}
\eea
We note that $M_{mn}\sim \lambda_{\nl}/R$ for all $m,n$, as expected for an IR
localized mass.
These Majorana masses mix\footnote{One could
instead include the boundary mass in
  the IR boundary conditions and obtain the full KK spectrum
  directly~\cite{Huber:2002gp}. Our main points will be readily
  seen treating the
  boundary terms as perturbations.}  the KK modes and the true mass
eigenstates are linear combinations of  $\nu^{(n)}$. As we will detail below, the spectrum
consists of a tower of Majorana neutrinos with masses starting at $\sim R^{-1}$.
For $\lambda_{\nl}\sim 0.1$ one has $M_{00}\sim  R^{-1}/10$ and the
lightest mode is predominantly composed
of $\nu^{(0)}_R$. The higher modes are pseudo-Dirac neutrinos with
mass splittings set by $M_{mn}< M_n$. The Dirac mass $M_n$ increases with $n$ as
$M_n\sim (n+c/2)\pi/R$ while $M_{mn}$ does not
significantly change, so the Majorana masses are increasingly
unimportant for the higher modes.

Having outlined the spectrum of sterile neutrinos we can proceed to
consider their coupling to the SM. This occurs via a UV localized
Yukawa interaction
\bea
S&\supset &-\frac{\lambda}{\sqrt{M_*}} \int d^5x\sqrt{-g_{uv}} \ \bar{L}
H N_{R}\ \delta(z-k^{-1})\ ,
\eea
where $L$ ($H$) is the SM lepton (scalar) doublet. After
integrating out the extra dimension one obtains Dirac mass terms
coupling the KK neutrinos to the SM:
\bea
S&\supset &-\sum_n  \int d^4x\ m_n^{\D}\  \bar{\nu}_L \nu^{(n)}_{R}\  ,
\eea
where $m^{\D}_n  = (\lambda \langle H\rangle /\sqrt{M_*})\times
f^{(n)}_R(k^{-1})$ and $\langle H\rangle\simeq 174$~GeV is the vacuum
value of the SM scalar. Using the approximations for
$f^{(n)}_R(k^{-1})$ with $c\simeq1$ gives 
\bea
m_0^{\D} &\simeq&  \lambda\ \sqrt{\frac{k(1+2c)}{M_*}} \
(kR)^{-c-1/2}\ \langle H\rangle\ ,\\
m_n^{\D} &\simeq&  2\lambda\ \sqrt{\frac{2}{M_*R}} \
\left(\frac{M_n}{2k}\right)^{c}\ \langle H\rangle\ . 
\eea

We will consider the full neutrino mass matrix in more detail and determine the spectrum of physical mass eigenstates in Section~\ref{sec:mass_eigenstates}. Presently, however, we wish to make a few pertinent comments. Including the boundary coupling to SM neutrinos produces a somewhat
complicated mass matrix describing both the SM and KK
neutrinos. Despite this, the
hierarchy of scales generated by the KK profiles allows the basic
spectrum to be readily understood. For $n>0$ the previously
pseudo-Dirac neutrinos now have Dirac masses coupling them to the SM,
which are roughly $(m^{\D}_n/M_n)\simeq \lambda
(\langle H\rangle/k) \sqrt{2/M_*R}\ll 1$ for $c\simeq1$.
This coupling can essentially be
treated as a perturbation  and to leading order the $n>0$ modes remain as
pseudo-Dirac neutrinos, comprised predominantly of sterile KK modes. The coupling between the SM and the zero mode is more important: to leading order these two essentially form a standard seesaw pair. The heavy
mode has mass of order $R^{-1}$ and is comprised mostly of $\nu^{(0)}$,
while the mass of the light SM neutrino is~\cite{McDonald:2010jm}
\bea
m_\nu\  \simeq\ \frac{(m^{\D}_0)^2}{M_{00}}\ \simeq\ \frac{\lambda^2}{\lambda_{\nl}}\ \frac{\langle
  H\rangle^2}{M_*}\ (kR)^{-2c}\ .\label{light_nu_mass}
\eea 
The key point is that light SM neutrino masses can be obtained
with relatively low values of the cutoff, $M_*\sim$~TeV,
due to the suppression induced by the extra factor $(kR)^{-2c}\ll1$.
No reliance on large UV scales is necessary. 

\begin{figure}[ttt]
\begin{center}
        \includegraphics[width = 0.7\textwidth]{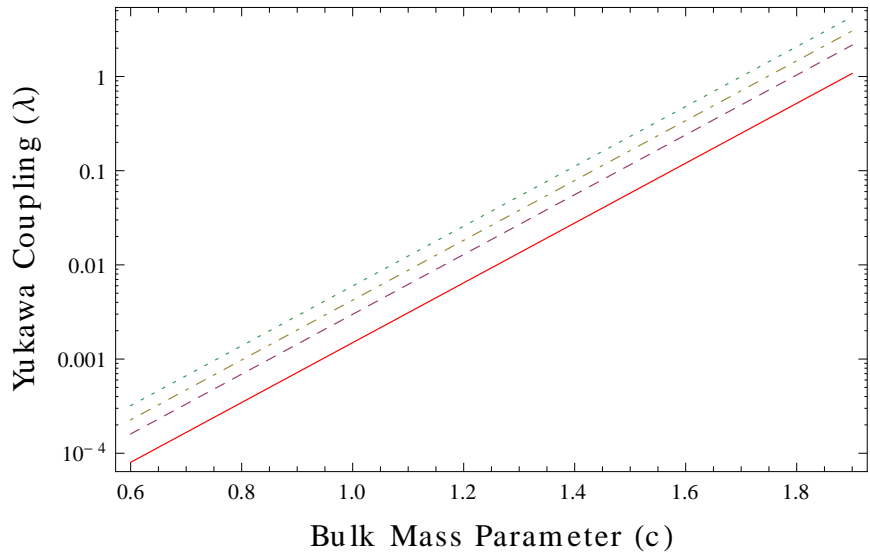}
\end{center}
\caption{The neutrino Yukawa Coupling ($\lambda$) as a function the Bulk
  Mass Parameter ($c$) for a fixed value of the SM neutrino
  mass ($m_\nu=0.1$~eV). The lines correspond to (bottom to top):
  $\lambda_{\nl}=0.1,\ 0.4,\ 0.8,\ 1.6$. The IR scale is fixed at $R^{-1}=1$~GeV
   and $k=M_*/2=1.5$~TeV.}
\label{fig:yuk_vs_c_normal}
\end{figure}

This approach, referred to as a ``mini-seesaw mechanism''~\cite{McDonald:2010jm}, generates naturally suppressed SM neutrino
masses in a two-fold process. Firstly, the effective 4D Dirac
and Majorana mass scales are suppressed: the sub-TeV Majorana
scale is generated by warping ($M_{00}\ll m_W$) while the Dirac mass
is suppressed by
a small wavefunction overlap ($m^{\D}_0\ll M_{00}\ll m_W$). Secondly,
a  low-scale seesaw operates
between the lightest KK mode and the SM neutrino, further
suppressing the SM neutrino mass. Together these
elements realize the order eV neutrino masses. We emphasize that,
unlike most seesaw models with light
sterile-Majorana scales, the small neutrino masses are generated naturally
and do not require tiny Yukawa couplings \emph{a priori}.  

Figure~\ref{fig:yuk_vs_c_normal} shows a plot of the Yukawa
coupling $\lambda$ as a function of the bulk mass
parameter\footnote{We restrict our attention to values of $|c|$ no
  larger than those employed in~\cite{Davoudiasl:2008hx}.} $c$ for
the fixed value $m_\nu=0.1$~eV. It is clear from the plot that $m_\nu$
can be readily suppressed below the weak 
scale to within the range of interest for the solar and atmospheric
neutrino data, even for $\mathcal{O}(1)$ values of the dimensionless
couplings $\lambda$ and $\lambda_{\nl}$. For future reference we note
that $\lambda\in[10^{-2},0.1]$ for $c\in[1.3,1.6]$  when
$\lambda_{\nl}=0.1$.

The goal of this work is to determine the experimental bounds on the
tower of low-scale sterile neutrinos. We will focus on values of
$R^{-1}\sim$~GeV and investigate whether the bounds allow
$\mathcal{O}(1)$ values for the Yukawa coupling $\lambda$. Before
turning to these matters, we note that, via the
application of the AdS/CFT
correspondence to RS models~\cite{ArkaniHamed:2000ds}, the present
model is considered dual to a 4D theory in which the UV localized SM fields are treated as
fundamental objects external to a hidden strongly-coupled sector. The
composite objects correspond to IR localized fields in the 5D picture
(roughly; see e.g.~\cite{Batell:2007jv}) and the lightest fermionic composite is the right-handed neutrino (or
neutrinos in a three generation model). Note that IR localization
of the Majorana mass term
indicates that lepton number is broken within the strongly coupled
sector.  The dual theory is conformal for energies
$\mathrm{TeV} \gtrsim E\gtrsim R^{-1}$, with the conformal symmetry broken
spontaneously in the IR (dual to the IR brane) and explicitly in the
UV (dual to the TeV brane, itself associated with the scale of electroweak symmetry
breaking). Thus the Little Warped Space allows one to model light right-handed neutrinos that are composites of a hidden CFT with a sub-weak mass gap. 

The small Yukawa coupling between the SM and the lightest right-chiral neutrino
is also understood in the dual picture. For every bulk field in the 5D picture, there exists a CFT
operator in the dual 4D theory that is sourced by a
fundamental field. The source field corresponds to
the UV-brane value of the given bulk field in the 5D theory. The SM is external to the CFT,
but couples directly to the source field, $N_R|_{z=1/k}$. Writing this field
in terms of the physical 
mass eigenstates introduces the aforementioned tiny mixing
angle, the dependence on which ensures the effective Yukawa coupling is highly
suppressed.
\section{Mixing with a Tower of Sterile Neutrinos\label{sec:mass_eigenstates}}
Having specified the model ingredients, we turn to a more detailed
consideration of mass mixing in the neutral lepton sector. The
neutrino mass Lagrangian can be written as:
\bea
\mathcal{L}_\nu\supset -\frac{1}{2}\ \bar{\mathcal{V}}^c_{flav}\  \mathcal{M}\ \mathcal{V}_{flav} \ +\ \mathrm{H.c.},
\eea
where the neutrino-space flavour basis vector is: 
\bea
\mathcal{V}_{flav} &=&(\nu_{\alpha L}, \nu_{0R}^{c}, \nu_{1R}^{c},
\nu_{1L},\nu_{2R}^{c}, \nu_{2L},\ \dots\ )^T\ ,
\eea
and the subscript ``$\alpha$'' denotes that this is a SM flavour eigenstate. The mass matrix is given by
\bea
\mathcal{M}=\left(\begin{array}{cc}0& \mathcal{M}_{D}^T\\\mathcal{M}_{D}& \mathcal{M}_H\end{array}\right)\ ,
\eea
where the Dirac mass matrix is
\bea
\mathcal{M}_{D}=(\ m_0^{\D},\ m_1^{\D},\ 0,\ m_2^{\D},\
0,\ \dots)^T\ ,
\eea
and the mass matrix for the ``heavy'' KK modes is
\bea
\mathcal{M}_{H}=
\left(\begin{array}{cccccc}
-M_{00}& -M_{01}&0& -M_{02}&0&\cdots\\
-M_{01}&-M_{11}&M_1& -M_{12}&0&\cdots\\ 
0&M_1&0&0&0&\cdots\\
-M_{02}& -M_{12}&0&-M_{22}&M_2&\cdots\\
0&0&0&M_2&0&\cdots\\
\vdots&\vdots&\vdots&\vdots&\vdots&\ddots
\end{array}\right)\ .
\eea
The Dirac masses $m_n^{\D}$ and the Majorana masses $M_{nn'}$ were defined in the previous section. A mass matrix with this form was discussed recently in Ref.~\cite{Watanabe:2009br} and an exact expression for the inverse was obtained:
\begin{eqnarray}
\mathcal{M}_H^{-1} 
\;=\; 
\left( \begin{array}{cccccc}
\frac{-1}{M_{{00}}} &0 & 
\!\!\!\frac{M_{{01}}}{M_{{00}}}\frac{-1}{M_{1}}\!\!\!
&0 & 
\!\!\!\frac{M_{{02}}}{M_{{00}}}\frac{-1}{M_{2}}\!\!
& \cdots \\
0&0 & \frac{1}{M_{1}} &0 &0 & \cdots \\
\!\!\frac{M_{{01}}}{M_{{00}}}\frac{-1}{M_{1}} & 
\frac{1}{M_{1}} &0 &0 &0 & \cdots \\
0&0 &0 &0 & \frac{1}{M_{2}} & \cdots \\
\!\!\frac{M_{{02}}}{M_{{00}}}\frac{-1}{M_{2}} &0 &0 &
\frac{1}{M_{2}} &0 & \cdots \\
\vdots & \vdots & \vdots & \vdots & \vdots & \ddots \\
\end{array}
\right)\ .
\label{inv}
\end{eqnarray}
We will use this result below. Note that, in deriving the above
expression for the inverse matrix, one makes use of relations between
the elements $M_{nn'}$ which result from the underlying warped space
derivation. Though the context in which this matrix was considered
in~\cite{Watanabe:2009br} differs from that considered here, the same
relations amongst the elements $M_{nn'}$ are present in our case. The
one generation result presented in Ref.~\cite{Watanabe:2009br} also
generalizes to three generations; i.e.~the elements of the matrix
presented in~\cite{Watanabe:2009br} (and here) can be treated as
matrices for the three generation case~\cite{watanabe}. We focus on
the one generation case but keep some expressions general. 

Diagonalization of the matrix $\mathcal{M}$ proceeds as follows. First we perform a standard seesaw rotation by defining the matrix
$\mathcal{U}$ as
\bea
\mathcal{U}\simeq\left(\begin{array}{cc}1& \mathcal{M}_{D}^T \mathcal{M}_H^{-1}\\-(\mathcal{M}_{D}^T \mathcal{M}_H^{-1})^T& 1\end{array}\right),
\eea
giving
\bea
\mathcal{U}^T\mathcal{M}\ \mathcal{U}\simeq \left(\begin{array}{cc}- \mathcal{M}_{D}^T \mathcal{M}_H^{-1}\mathcal{M}_{D}&0\\0& \mathcal{M}_H\end{array}\right),
\eea
in the basis $\mathcal{U}^\dagger  \mathcal{V}_{flav}$. To diagonalize
the heavy KK matrix we write it as
$\mathcal{M}_{H}=\mathcal{M}_{KK}+\mathcal{M}'$, where the first
(second) term contains the KK Dirac (Majorana) masses $M_n$ $(M_{nn'})$. We treat the Majorana masses as a perturbation on the matrix $\mathcal{M}_{KK}$. As shown in the previous section, the elements $M_{nn'}$ are all $\mathcal{O}(\lambda_{\nl}/R)$, while the KK masses grow like $M_n\sim n\pi/R$. Thus for $|\lambda_{\nl}|\lesssim1$ one has $|M_{nn'}|\ll M_n$ for $n\gg1$, and the Majorana masses merely perturb the spectrum of the higher KK modes. For the modes with $n\sim1$ one still has $|M_{nn'}|<M_n$ for $|\lambda_{\nl}|\lesssim1$, but the Majorana masses can be more important as $\lambda_{\nl}$ approaches and exceeds unity from below. For $\lambda_{\nl}\lesssim 10^{-1}$ it is valid to treat the Majorana masses as perturbations for the entire tower. For simplicity we focus on this region of parameter space but mention some differences where important.

Next we define the matrix $\tilde{\mathcal{U}}$ as\footnote{In a three
  generation analysis the entry ``$1$'' in (\ref{eq:rotation}) is
  replaced by the PMNS mixing matrix $U_\nu$, which diagonalizes the light-sector matrix $\mathcal{M}_\nu=- \mathcal{M}_{D}^T \mathcal{M}_H^{-1}\mathcal{M}_{D}$.}
\bea
\tilde{\mathcal{U}}=\mathrm{diag}\left(1, U_{KK}\right),\label{eq:rotation}
\eea
where $U_{KK}$ diagonalizes the KK matrix $\mathcal{M}_H$:
\bea
U_{KK}^T\mathcal{M}_{H}\ U_{KK} =
\mathrm{diag}(\ M_0,\ M_{+1},\ M_{-1},\ M_{+2},\ M_{-2},\ \dots)\ ,
\eea
and the leading order expression for $U_{KK}$ can be found in Appendix~\ref{sec:mass_mixing}. The mass eigenvalues are
\beq
m_\nu&\simeq& \frac{(m_0^{\D})^2}{M_{00}}\simeq\ \frac{\lambda^2}{\lambda_{\nl}}\ \frac{\langle
  H\rangle^2}{M_*}\ (kR)^{-2c}\nonumber\\
M_{0}&\simeq&-M_{00}\nonumber\\
M_{\pm n}&\simeq&\pm M_n - \frac{1}{2}M_{nn},
\eeq
where we include the light mass eigenvalue $m_\nu$ for
completeness. Negative mass eigenvalues can be brought positive by an
appropriate field redefinition, as per usual. We will leave the
negative signs so we don't need to keep track of phases in the mixing
matrix. Let us emphasize that the above expressions for the $n>0$ KK
masses are correct to $\mathcal{O}(M_{nn'}/M_n)$; their precise form
can change if $M_{nn'}$ is increased such that $M_{nn'}\gtrsim M_n$
for some values of $n$. However, the expression for the SM neutrino
mass, $m_\nu\simeq (m_0^{\D})^2/M_{00}$, is exact to
$\mathcal{O}(\mathcal{M}_{D}^T \mathcal{M}_H^{-1})$ and does not
depend on the relation between $M_{nn'}$ and $M_n$. It is
interesting that the leading order seesaw-suppressed mass
depends only on the Majorana mass of the zero-mode.

 The flavour basis can be expressed in terms of the mass basis:
\bea
\mathcal{V}_{flav} &=&\mathcal{U} \tilde{\mathcal{U}} \mathcal{V}_{mass}\nonumber\\
&\simeq& \left(\begin{array}{cc}1&(\mathcal{M}_{D}^T \mathcal{M}_H^{-1}) U_{KK}\\-(\mathcal{M}_{D}^T \mathcal{M}_H^{-1})^T &U_{KK}\end{array}\right) \mathcal{V}_{mass}\ ,
\eea
where the  mass eigenstates are written as
\bea
\mathcal{V}_{mass} &=&(\nu_{i L}, N_{0R}^{c}, N_{+1L},
N_{-1L},N_{+2L}, N_{-2L},\ ...\ )^T\ ,
\eea
and the subscript ``$i$'' denotes a (mostly) SM mass eigenstate. We can write the SM flavour eigenstate in terms of the mass
eigenstates as
\beq
\nu_{\alpha L}= \nu_{iL}+\theta_{\alpha 0}N_{0R}^c +\sum_{n>0}\left(\theta_{\alpha,+n} N_{+n}  +\theta_{\alpha,-n} N_{-n}  \right)_L,
\eeq
where $\theta_{\alpha, \pm n}$ is the mixing matrix element between
$\nu_{\alpha}$ and the $\pm n$th
level KK mode.  For much of what follows we will not need to
differentiate the SM flavour eigenstates and accordingly refer
 to $\theta_{\alpha, \pm n}$ simply as $\theta_{ \pm n}$. In
 the limit of a vanishing boundary Majorana mass,
 $\lambda_{\nl}\rightarrow 0$, the two modes $N_{\pm n}$ fuse
together to form a single Dirac neutrino with mass $M_n$, and one has
$N_{\pm n L}=(\nu_{n R}^c \pm \nu_{nL})/\sqrt{2}$.

Treating the boundary Majorana mass terms as a perturbation on the KK
 mass matrix, the mixing angles are approximately
\bea
\theta_{0}&\simeq& -\frac{m_{0}^\D}{M_{00}} +\sum_{n>0}\frac{M_{0n}}{M_n}\frac{m_n^\D}{M_n}\nonumber\\
\theta_{\pm n}&\simeq&\frac{1}{\sqrt{2}}\frac{m_n^\D}{M_n}\left(\pm
  1+\frac{M_{nn}}{4M_{n}}\right)
+\sum_{n'\ne n}\frac{m_{n'}^\D}{\sqrt{2}} \left(\frac{M_{nn'}}{M_{n'}^2-M_n^2}\right).\nonumber
\eeq
We will make use of these expressions in what follows. Note that
 the $n$-dependence of the Majorana mass appearing in the sum term of
 $\theta_0$ is $M_{0n}\sim (-1)^n/R$. Consequently the
sum piece in the expression for $\theta_0$ is found to be subdominant for
$\lambda_{\nl}\lesssim 0.1$, giving $\theta_0\simeq -m_0^\D/M_{00}$,
which is of the usual seesaw form. Similarly, the sum piece in the
expressions for $\theta_{\pm n}$ gives a sub-leading contribution.

Before turning to a detailed analysis of the bounds, we note that the hidden sector also contains a tower of KK gravitons with mass
splittings of order $R^{-1}\sim$~GeV. In the present work we follow
Ref.~\cite{Davoudiasl:2008hx}  and give the bulk graviton a Dirichlet
BC on the UV brane. This projects the zero-mode graviton out of the
spectrum and ensures that the higher modes do not couple directly to
the UV-localized SM. In this case 4D gravity is not included and the
model deals purely with particle physics. The spectrum also contains a
physical 4D scalar (the radion) that can couple to the UV-localized
SM. However, with the standard parametrization of the radion
fluctuation~\cite{Charmousis:1999rg} we find that the radion $r(x)$
couples to the SM fields like $r(x)/\Lambda_r$, with the dimensionful
parameter being $\Lambda_r\sim (kR)\times  M_*\gg M_*$. Radion interactions
can thus be safely neglected for $M_*\sim$~TeV. In a more complete
treatment, one could aim to retain 4D gravity
by including a large
UV-localized Einstein-Hilbert term and analyzing the metric
fluctuations in the interval approach to braneworld
gravity~\cite{Carena:2005gq}. The KK gravitons would be largely
repelled from the UV brane so their coupling to UV-localized SM fields
remains sufficiently weak to avoid experimental constraints. The radion coupling may also be modified by
this procedure. We hope to study these matters in a future work. 
\section{Neutrinoless Double Beta-Decay\label{sec:nubb}}
With the above information we have all we need to study the bounds on
the mini-seesaw. The sterile KK modes provide additional channels for neutrinoless
double beta-decay ($\nubb$-decay), the non-observation of
which provides bounds on the parameter space.  These bounds can be approximated by noting that the
amplitude for $\nubb$-decay from SM neutrinos goes like~\cite{Rodejohann:2010wi}
\bea
\mathcal{A}\ \propto\  G_F^2\ \frac{\langle m_{ee}\rangle}{q^2}\
,
\eea 
where $\langle m_{ee}\rangle= \sum_i|U_{ei}^2m_i|$, $G_F$ is the Fermi
constant, and $q^2\simeq (0.1\ \mathrm{GeV})^2$ is the typical momentum exchanged during the process. The amplitude for $\nubb$-decay
induced by a single sterile Majorana neutrino ($N_s$) that mixes with the SM
neutrino via $\nu_\alpha \simeq \nu_i +\theta_s N_s $ goes like
\bea
\mathcal{A}_s\ \propto\  G_F^2\ \frac{\theta_s^2}{M_s}\ .
\eea
Using the bound $\langle m_{ee}\rangle\lesssim 0.5\ \mathrm{eV}$ and
assuming that the sterile neutrino contribution dominates, one obtains
the
bound $\theta_s^2/M_s\lesssim 5\times 10^{-8}\ \mathrm{GeV}^{-1}$~\cite{Rodejohann:2010wi,KlapdorKleingrothaus:1999kq}.

We can generalize this result for a tower of sterile neutrinos. The
$\nubb$-decay amplitude produced by the tower is 
\bea
\mathcal{A}_{KK}\propto \
G_F^2\ \frac{\theta_0^2}{M_0}+\sum_{n>0}\ G_F^2\ \left\{\frac{\theta_{-n}^2}{M_{-n}}+
  \frac{\theta_{+n}^2}{M_{+n}}\right\},
\eea
and one therefore requires
\bea
3\times \left|\frac{\theta_0^2}{M_0}+\sum_{n>0}\ \left\{\frac{\theta_{-n}^2}{M_{-n}}+
  \frac{\theta_{+n}^2}{M_{+n}}\right\}\right|\ \lesssim\  5\times
10^{-8}\ \mathrm{GeV}^{-1},
\eea
where we make use of the fact that $M_n^2\gg q^2$ for the parameter
space we consider. We also include a sum over  three generations of sterile
neutrinos, assuming that the active/sterile mixing is similar for each
generation (i.e.~we do not assume any tuned cancellations of leading
terms~\cite{de Gouvea:2007uz,Asaka:2011pb}). With the information from the previous sections one
can calculate the the above factor and determine the size of
the $\nubb$-decay amplitude. In general the resulting expressions are
rather complicated and numerical evaluation is required. However, some
simple leading-order expressions can be obtained for the
regions of parameter space considered here. Approximating the mixing angles by their leading terms:
\bea
\theta_{0}&\simeq& -\frac{m_{0}^\D}{M_{00}},\nonumber\\
\theta_{\pm n}&\simeq& \frac{1}{\sqrt{2}}\frac{m_n^\D}{M_n}\left(\pm 1+\frac{M_{nn}}{4M_{n}}\right),\nonumber
\eeq
and noting that
\bea
\sum_{n>0}\ \left\{\frac{\theta_{-n}^2}{M_{-n}}+
  \frac{\theta_{+n}^2}{M_{+n}}\right\}\simeq \sum_{n>0}\ \frac{1}{M_{n}}\left\{\theta_{+n}^2 -\theta_{-n}^2 +\frac{M_{nn}}{2M_{n}}(\theta_{+n}^2 +\theta_{-n}^2 ) +\ \dots \right\},
\eea
one has
\bea
\sum_{n>0}\ \left\{\frac{\theta_{-n}^2}{M_{-n}}+
  \frac{\theta_{+n}^2}{M_{+n}}\right\}\simeq \sum_{n>0}\ \left(\frac{m_n^\D}{M_n}\right)^2\left(\frac{M_{nn}}{M_{n}}\right)\frac{1}{M_{n}}+\ \dots\ .
\eea
Note that the $n$th term is proportional to the lepton number
violating mass $M_{nn}$; in the limit that the Majorana mass vanishes
the contributions to $\nubb$-decay from the modes $\pm n$ cancel each
other out. For $c$ in the vicinity of unity the above may be expressed in terms of the
light SM neutrino mass:\footnote{We have taken $\lambda_{\nl}>0$ in
  Eq.~(\ref{nubb}). More generally, the factor of ``$-1$''
   is replaced by $(-1)\times \mathrm{sgn}(\lambda_{\nl})$.} 
\bea
& &3\times\left[\frac{\theta_0^2}{M_0}+\sum_{n>0}\ \left\{\frac{\theta_{-n}^2}{M_{-n}}+
  \frac{\theta_{+n}^2}{M_{+n}}\right\}\right]\label{nubb}\\
&\simeq&
\left(\frac{m_\nu}{10^{-1}\ \mathrm{eV}}\right)
\left(\frac{10^{-1}}{\lambda_{\nl}}\right)^2
\left(\frac{\mathrm{GeV}}{R^{-1}}\right)^2\times
\left|\frac{-3}{(1+2c)^2} + 3\lambda_{\nl}^4\sum_{n>0}
  \left(\frac{M_n R}{2}\right)^{2(c-2)}\right|\times 10^{-8}\ \mathrm{GeV}^{-1}.\nonumber
\eea
The above expression can be evaluated numerically, with the
sum over KK modes cutoff in the UV at $n_*\simeq M_*R/\pi$. In
practise, no significant error is introduced if one takes the RS2
limit and replaces the sum by a definite integral: $\sum_{n=1}^{n_*}\rightarrow \int_1^{n_*}dn$.  

\begin{figure}[ttt]
\begin{center}
        \includegraphics[width = 0.75\textwidth]{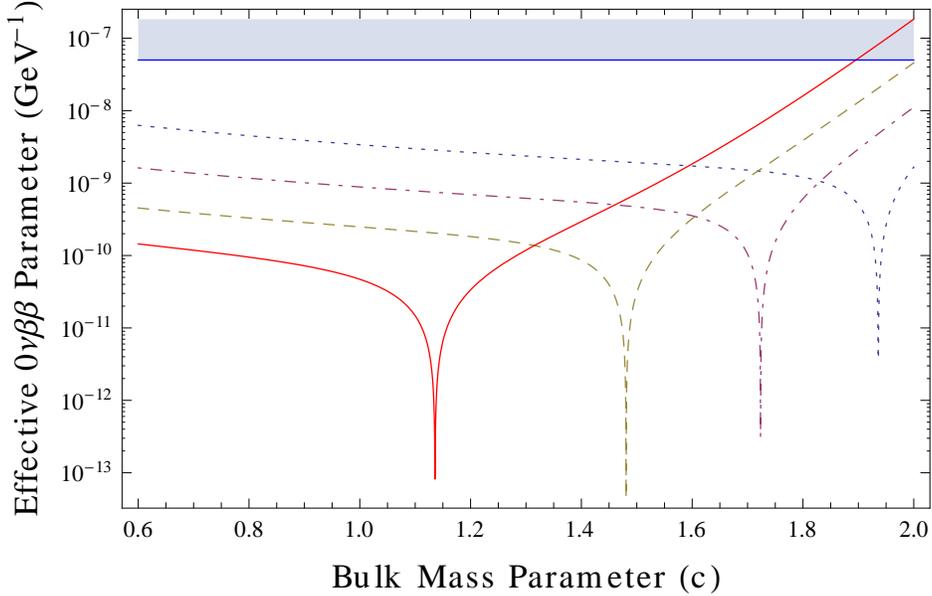}
\end{center}
\caption{The Effective Neutrinoless Double Beta-Decay Parameter
  ($\sum_n\theta_n^2/M_n$) as a function of the Bulk
  Mass Parameter ($c$). The SM neutrino masses are fixed at
  $m_\nu=0.1$~eV and the dotted (dot-dashed, dashed, solid) line corresponds to
  $\lambda_{\nl}=0.1\ (0.2,\ 0.4,\ 0.8)$. The shaded region is
  experimentally excluded. We use
  $R^{-1}=1$~GeV and
  $k=M_*/2=1.5$~TeV for the plot.}
\label{fig:nubb_vs_c}
\end{figure}

We plot Eq.~(\ref{nubb}) as a function
of the bulk mass parameter $c$ 
for a fixed value of the SM neutrino mass ($m_\nu=0.1$~eV) in
Figure~\ref{fig:nubb_vs_c}. For completeness we include the sub-leading terms in the mixing angles for the plot, though the result is well approximated by the above expression. The experimental bound is included in the
plot for comparison and we take
$R^{-1}=1$~GeV, $k=1.5$~TeV and $k/M_*=1/2$. Observe that the effective
$\nubb$-decay parameter is below the experimental bound for most of
the parameter space: all but the largest values of $c$ are viable, with some dependence on $\lambda_{\nl}$.

It is instructive to comment on the $\lambda_{\nl}$ dependence in the
above expressions (and the figure). For small values of $\lambda_{\nl}$
the amplitude for $\nubb$-decay is dominated by the zero-mode
contribution. This is because the KK neutrinos $N_{\pm n}$ form a
pseudo-Dirac pair for small $\lambda_{\nl}$ and the $\nubb$-decay
contributions from the two modes tend to cancel each other out. The
zero mode has no pseudo-Dirac partner and for small $\lambda_{\nl}$
its mass is the only one that ``badly'' breaks lepton number symmetry;
thus its contribution to the lepton-number violating $\nubb$-decay
rate is dominant. For small values of $\lambda_{\nl}$ the overall
$\nubb$-decay rate diminishes with increasing $\lambda_{\nl}$ due to
the increase in
the zero-mode mass. This behaviour
is observed in the small $c$ regions in Figure~\ref{fig:nubb_vs_c}.

On the other hand, for larger values of $\lambda_{\nl}$ the
contribution from the KK modes is seen to dominate. The mass splitting
between the pseudo-Dirac KK pairs increases for increasing $\lambda_{\nl}$, $|M_{-n}|-|M_{+n}|\sim \lambda_{\nl}/ R$.  The KK modes
$N_{\pm n}$ thus become less pseudo-Dirac-like and form a standard
pair of Majorana particles. Accordingly, lepton-number symmetry is
more severely broken in the higher KK sector and the $\nubb$-decay
signal increases. The cusps in Figure~\ref{fig:nubb_vs_c} occur at the transition
between zero-mode dominance and KK mode dominance; one
observes that, once the KK contribution is dominant, the overall rate
increases with $\lambda_{\nl}$ for a given value of $c$.

For perturbative values of $\lambda_{\nl}$, only the
low-lying modes ($n\lesssim 10$) are expected to have
sizeable mass splittings and thus contribute significantly to $\nubb$-decay. The relative splitting goes like $1/n$,
$(|M_{-n}|-|M_{+n}|)/M_n\sim \lambda_{\nl} /n\pi$, and the decreased
splitting suppresses the contribution of the higher modes. 

We have considered the KK neutrino contribution to $\nubb$-decay in
this section. There is also a contribution from the SM
neutrinos which, in general, will interfere with the
zero-mode contribution (these two form a standard seesaw pair and
thus possess opposite CP parities). However, the contributions from the
$n>0$ modes and the SM neutrinos can add constructively, so the
total rate can be larger if $c$ is such that the KK contribution
dominates the zero-mode. Given the potential experimental reach for
 an inverted mass hierarchy~\cite{Dueck:2011hu}, the additional
 contribution from
the KK neutrinos can improve the prospects for $\nubb$-decay
observation. Let us emphasize, however, that the SM contribution will
typically dominate the KK contribution in the region of parameter
space where the largest KK contribution comes from the zero-mode. It
has been argued that, generically, in seesaw models the contribution
from the SM neutrinos is larger than the sterile neutrino
contribution~\cite{Blennow:2010th}. This holds in the present case for
the $\nubb$-decay contributions from the zero-mode and the SM neutrino, which essentially form a standard
seesaw pair. Indeed, for $\langle m_{ee}\rangle\sim m_\nu=0.1$~eV one obtains
a contribution of $\langle m_{ee}\rangle/q^2\sim
10^{-8}~\mathrm{GeV}^{-1}$ to the quantity plotted in
Figure~\ref{fig:nubb_vs_c}, which always dominates the zero-mode
contribution, but can be exceeded by the sum over higher KK modes for
larger $c$. We only plot the KK contribution in
Figure~\ref{fig:nubb_vs_c} to emphasize the behaviour of this new contribution. 
\section{Invisible Z-Decays\label{sec:hid_z_decay}}
Mixing between the SM neutrinos and sterile neutrinos induces an
effective coupling between the $Z$ boson and the sterile neutrinos; if
the sterile neutrinos are light enough, additional invisible $Z$-decay
channels exist~\cite{Gronau:1984ct,Dittmar:1989yg}. When a heavy
sterile neutrino with mass $m_N< m_Z$ mixes with the SM neutrinos,
$\nu_i\simeq \nu_\alpha +\theta_s N$ with $\theta_s\ll1$, one has
\bea
\Gamma(Z\rightarrow \nu N)=\Gamma(Z\rightarrow 2\nu)\ |\theta_s|^2\ \left[1-\frac{m_N^2}{m_Z^2}\right]^2\left[1+\frac{m_N^2}{2m_Z^2}\right].
\eea
This result generalizes in an obvious way when a tower of KK neutrinos
mixes with the SM:
\bea
\Gamma(Z\rightarrow \nu N_{\pm n})=\Gamma(Z\rightarrow 2\nu)\
|\theta_{\pm n}|^2\ \left[1-\frac{M_{\pm n}^2}{m_Z^2}\right]^2\left[1+\frac{M_{\pm n}^2}{2m_Z^2}\right].\label{z_decay_width_one_kk}
\eea
Note that $\Gamma(Z\rightarrow N_{n'}N_{
  n})\ll\Gamma(Z\rightarrow \nu N_{ n})$ for $|\theta_n|^2\ll1$ and
one need only consider single production. The total non-standard contribution to the invisible width is thus
\bea
\Gamma(Z\rightarrow \nu+\mathrm{invisible})=9\times\sum_n\Gamma(Z\rightarrow \nu
N_{\pm n}),
\eea
where three families of active and sterile neutrinos are included, and
we again ignore the mixing of the three generations. Neglecting the final state KK neutrino mass in
(\ref{z_decay_width_one_kk}), this is approximately 
\bea
\Gamma(Z\rightarrow
\nu+\mathrm{invisible})\simeq9\ \Gamma(Z\rightarrow 2\nu)\times\left[|\theta_0|^2+\sum_{n>0}\left(|\theta_{+n}|^2+|\theta_{-n}|^2\right)\right].
\eea

\begin{figure}[ttt]
\begin{center}
        \includegraphics[width = 0.7\textwidth]{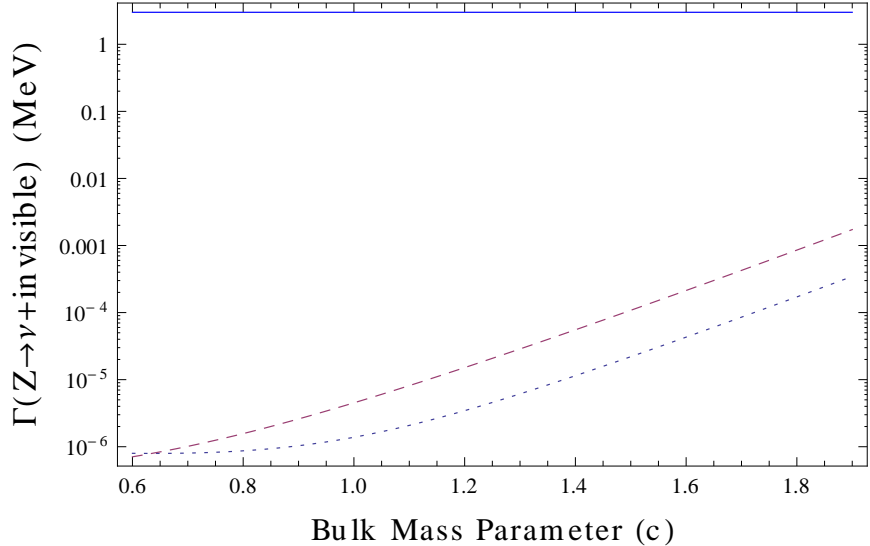}
\end{center}
\caption{The Invisible $Z$-Width as a function of the Bulk
  Mass Parameter ($c$) for fixed values of the SM neutrino
  masses, $m_\nu=0.1$~eV. The dashed (dotted) line is for
  $\lambda_{\nl}=0.5\ (0.1)$ and the solid line at the top shows the
  upper bound. We use
  $R^{-1}=1$~GeV,
  $k=1.5$~TeV and $k/M_*=1/2$ for the plot.}
\label{fig:Invis_Z_width_vs_c}
\end{figure}

The invisible $Z$-width has been precisely measured as
$\Gamma(Z\rightarrow\mathrm{invisible})=499.0\pm1.5$~MeV~\cite{PDG:2010}. This
should be compared to the SM prediction of $501.65\pm0.11$~MeV for $Z$
decays to neutrinos. New contributions are severely constrained and
should essentially lie within the experimental uncertainty, the
$2\sigma$ value for which is of order MeV. We plot the invisible
$Z$-width as a function of the bulk mass parameter $c$ in
Figure~\ref{fig:Invis_Z_width_vs_c}. The SM neutrino mass is fixed at
$m_\nu=0.1$~eV and we take
$R^{-1}=1$~GeV, $k=1.5$~TeV and $k/M_*=1/2$, while varying $\lambda_{\nl}$. One observes that the invisible width is well below the
$O(1)$~MeV experimental bound for the entire region of parameter space.

\section{Other Bounds\label{sec:other_bounds}}
Analysis of the bounds on a single sterile-neutrino with
$\mathcal{O}(\mathrm{GeV})$ mass can be found in~\cite{Smirnov:2006bu}
and, more recently, in~\cite{delAguila:2008pw,Atre:2009rg}. The lightest (zero mode)
neutrinos in the present model must satisfy these standard bounds. The
most robust constraints on an order $100$~MeV sterile neutrino come
from peak searches in the leptonic decays of pions and
kaons~\cite{Britton:1992pg,Berghofer:1981ty}. These constraints
assume only
that the heavy neutrino mixes with the SM  and are therefore insensitive to the decay
properties of the neutrinos. This differs from most collider bounds, which
typically make assumptions about the neutrino decay properties to
define the signal. For $R^{-1}\sim$~GeV only the zero mode neutrinos can
appear in the pion/kaon decays and one may use the data to constrain $ |\theta_0|^2$, which may itself be cast as:
\bea
  |\theta_0|^2\sim  \frac{10^{-9}}{3}\times \left(\frac{m_\nu}{0.1~\mathrm{eV}}\right)\left(\frac{\mathrm{GeV}}{R^{-1}}\right)\left(\frac{10^{-1}}{|\lambda_{\nl}|}\right),
\eea
One observes that the constraints of $ |\theta_0|^2\lesssim10^{-6}$ obtained from kaon decays and $ |\theta_0|^2\lesssim10^{-5}-10^{-8}$ from pion decays (see Figure~3 in Ref.~\cite{Atre:2009rg}) are not severe.

 Ref.~\cite{Nardi:1994iv} has detailed a number of bounds for sterile neutrinos
that mix with SM neutrinos. Their focus was primarily on neutrinos
with mass $M_N>m_Z/2$ but some of their bounds are reliable for
lighter masses. Noting that mixing with sterile neutrinos in general
introduces a correction to the muon decay constant they derive bounds from
lepton universality, CKM unitarity\footnote{The observed CKM mixing elements are
extracted with some sensitivity to the muon decay constant; a shift in
the latter modifies the former.} and the invisible $Z$-decay width. For
 $M_N>m_Z/2$ the latter is affected by the (slightly)
modified coupling of SM neutrinos to the $Z$; in this work we
have KK neutrinos satisfying $M_N< m_Z/2$ and the invisible $Z$ width
is modified directly by the additional decay channels $Z\rightarrow
\nu N_n$ as detailed in Section~\ref{sec:hid_z_decay}. 

The bounds of
Ref.~\cite{Nardi:1994iv} can be
summarized as
\bea
 \theta_e^2&<&7.1\times10^{-3}\quad \mathrm{for}\quad
M_N>0.14~\mathrm{GeV}\ (m_\pi),\nonumber\\
 \theta_\mu^2&<&1.4\times10^{-3}\quad \mathrm{for}\quad
M_N>1.15~\mathrm{GeV}\ (m_\Lambda),\nonumber\\
 \theta_\tau^2&<&1.7\times10^{-2}\quad \mathrm{for}\quad
M_N>1.777~\mathrm{GeV}\ (m_\tau).\label{sterile_nardi_bounds}
\eea 
Here $\theta_\alpha^2$ is an effective mixing angle between the
SM neutrino with lepton flavour $\alpha$ and sterile neutrinos with
masses exceeding the quoted values. In the present context the
effective angle is defined
as
$\theta_{\mathrm{eff}}^2/3=\theta_0^2+\sum_n(\theta_{-n}^2+\theta_{+n}^2)$,
with the factor of three accounting for three generations of bulk neutrinos. For
simplicity we do not differentiate the different
generations in this work and assume that all mixing angles are of
the same order. In this case a conservative bound of
\bea
\theta_{\mathrm{eff}}^2=3\times\left\{\theta_0^2+\sum_{n>0}(\theta_{-n}^2+\theta_{+n}^2)\right\}&<&7.1\times10^{-3}\quad \mathrm{for}\quad
M_0>\ m_\pi,
\eea
is obtained. We plot the effective active-sterile mixing-angle squared
as a function of the bulk mass parameter ($c$) in
Figure~\ref{fig:mixing_vs_c}, along with the experimental bound. The figure shows that values of
$c\lesssim1.85\ (1.7)$ are consistent with the bounds for
$\lambda_{\nl}=0.1\ (0.5)$.

\begin{figure}[ttt]
\begin{center}
        \includegraphics[width = 0.7\textwidth]{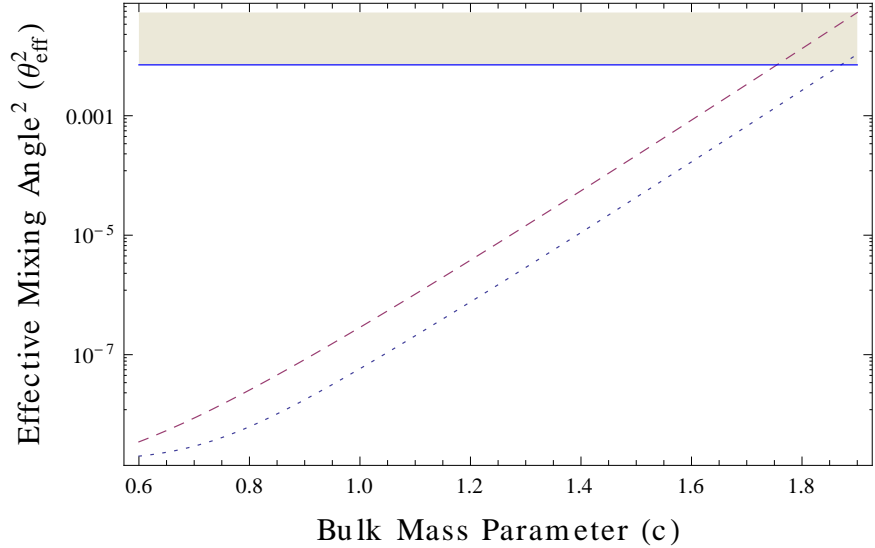}
\end{center}
\caption{The Effective Active-Sterile Mixing-Angle Squared
  ($\sum_n\theta_n^2$) as a function of the Bulk
  Mass Parameter ($c$) for fixed values of the SM neutrino
  masses, $m_\nu=0.1$~eV. The dashed (dotted) line is for
  $\lambda_{\nl}=0.5\ (0.1)$ and the shaded region is excluded. We use
  $R^{-1}=1$~GeV,
  $k=1.5$~TeV and $k/M_*=1/2$ for the plot.}
\label{fig:mixing_vs_c}
\end{figure}

Light sterile neutrinos can also be created in collider
experiments. Kinematics permitting, sterile neutrinos can
be singly produced 
on the $Z$ peak and can decay via an off-shell gauge boson, 
$N\rightarrow \nu +Z^*$ with
$Z^*\rightarrow \ell^+\ell^-,\ \bar{q}q$, or $N\rightarrow \ell +W^*$
with $W^*\rightarrow \ell\nu_{\ell},\ qq'$. These decay signals have been
sought at LEP for gauge-singlet neutrinos in the mass range
$3~\mathrm{GeV}<M_N<m_Z$~\cite{Adriani:1992pq}. The resulting bound is
 essentially constant at $|\theta_{s}|^2\lesssim 2\times 10^{-4}$ for $M_N$ in
 the range $3~\mathrm{GeV}<M_N<50~\mathrm{GeV}$ and decreases slowly (rapidly) for
 lighter (heavier) $M_N$; see Figures~6 and~7 in
 Ref.~\cite{Adriani:1992pq}. 

In the present framework,  each member of the KK tower can decay to SM
fields via off-shell gauge bosons\footnote{Modes with mass $M_n>m_Z$
  can also decay directly to on-shell vectors.} and it appears that
the LEP bounds
should be applied to all of the KK neutrinos. The LEP bounds certainly apply
to the zero mode neutrinos, which decay through their inherited SM
couplings and 
behave essentially like a standard, light, sterile neutrino. One thus
requires $|\theta_{0}|^2\lesssim 10^{-4}$ for $(M_0/\mathrm{GeV})\in [3,50]$, with the bound dropping to
$|\theta_{0}|^2\lesssim 10^{-2}$ as $M_0$ decreases to $500$~MeV. Given that $\theta_0^2\simeq 10^{-10}\times (m_\nu/10^{-1}\mathrm{eV})\times(\mathrm{GeV}/|M_{00}|)$ this bound is not restrictive.

The higher  modes ($n>0$) can also decay to the SM. However, one should not
apply the bounds of Ref.~\cite{Adriani:1992pq} directly to these modes.
These bounds are obtained under the assumption that the sterile
neutrinos \emph{only} decay to
Standard Model final states. This assumption does not hold for
the $n>0$ KK modes, which can decay within the hidden sector via KK
graviton production,
$N_n\rightarrow N_{m}+h_a$, where $h_a$ is the $a$th KK graviton. The KK
neutrinos and gravitons are both localized towards the IR brane and in
general have large wavefunction overlaps. Despite the vertex
being gravitational in nature, the relevant coupling
strength is of order $R\gg M_{Pl}^{-1}$ and the hidden sector decays cannot be
neglected.  Graviton decays are
allowed for $n>m+a$ and the partial widths go like
$\Gamma_n\sim (k/M_*)^3 M_n^3R^2$. The modes are  increasingly broad as one
goes up the KK tower and, for a given value of $n$, decays with $n\sim
a+m$ are preferred. The total two-body hidden width for the $n$th mode
is thus of order\footnote{This behaviour is similar to the hidden KK vector widths
  detailed in Ref.~\cite{McDonald:2010fe}.}
$ n\times \Gamma_n$. Decay of the $n$th mode to the SM is suppressed by a factor of $|\theta_{\pm n}|^2\ll 1$ so that, absent hierarchically small values of
$k/M_*$, decays within the hidden sector will
dominate. 

Once a given KK neutrino decays into the hidden
sector, the daughter neutrino(s) and graviton(s) will further decay
within the hidden sector (when kinematically allowed). The graviton can
decay via $h_a\rightarrow
N_{m}N_{n}$ with width $\Gamma_a\sim (k/M_*)^3 m_a^3 R^2$ (similar
to RS~\cite{Davoudiasl:1999jd}) and shows a preference for decays with
$a\sim m+n$. Thus a hidden sector ``shower'' occurs,
consisting of a series of cascade decays down the hidden KK towers until one
has a collection of 
light KK modes with mass of order $R^{-1}$, which in turn decay back
to the SM. The 
expected signal from single production of a higher KK neutrino is
therefore very
different to the signal searched for at LEP~\cite{Adriani:1992pq}.
Rather than applying the constraint $|\theta_{s}|^2\lesssim
10^{-4}$ directly to $|\theta_{\pm n}|$ one should instead apply it to the
quantity $BR(N_{\pm n}\rightarrow \mathrm{SM})\times
|\theta_{\pm n}|^2$.  Given that
$BR(N_{\pm n}\rightarrow
\mathrm{SM})\ll 1$ and $|\theta_{\pm n}|^2<10^{-4}$ for the parameters of interest here ($k\lesssim M_*\sim$~TeV, $R^{-1}\sim$~GeV and $c\simeq1$), the LEP bound does
not impose an additional constraint. Bounds resulting from heavy neutrino searches at beam dump
experiments by, e.g., the WA66
Collaboration~\cite{CooperSarkar:1985nh}, are similarly unrestrictive
in the present context.


\section{FCNC Bounds: $\boldsymbol{\mu\rightarrow e+\gamma}$\label{sec:mu_to_e_gamma}}
Indirect constraints on the parameter space for sterile neutrinos can be
derived from searches for flavour changing neutral current processes
like $\mu\rightarrow e+e^+e^-$, $\mu\rightarrow e+\gamma$ and $\mu-e$
conversion in nuclei~\cite{Tommasini:1995ii}. For sterile neutrinos
mixing with the SM, the induced $\mu\rightarrow e+\gamma$ branching fraction
can be written as~\cite{Tommasini:1995ii,Ma:1980gm,Langacker:1988ur}:
\bea
BR(\mu\rightarrow e+ \gamma)= \frac{3\alpha}{8\pi}\ \left|\sum_n
  \theta_{en}\ \theta^*_{\mu n}\ \mathcal{F}(M_n^2/m_W^2)\right|^2,
\eea
where the function $\mathcal{F}(x)$ is
\bea
\mathcal{F}(x)=x[1-6x+3x^2+2x^3-6x^2\ln(x)]/2(1-x)^4,
\eea
with $\mathcal{F}(x)\in[0,1]$ for $x\in[0,\infty)$. Present
bounds require $BR(\mu\rightarrow e+ \gamma)<1.2\times 10^{-11}$ at
the $90\%$ C.L.~\cite{Brooks:1999pu}. 
\begin{figure}[ttt]
\begin{center}
        \includegraphics[width = 0.7\textwidth]{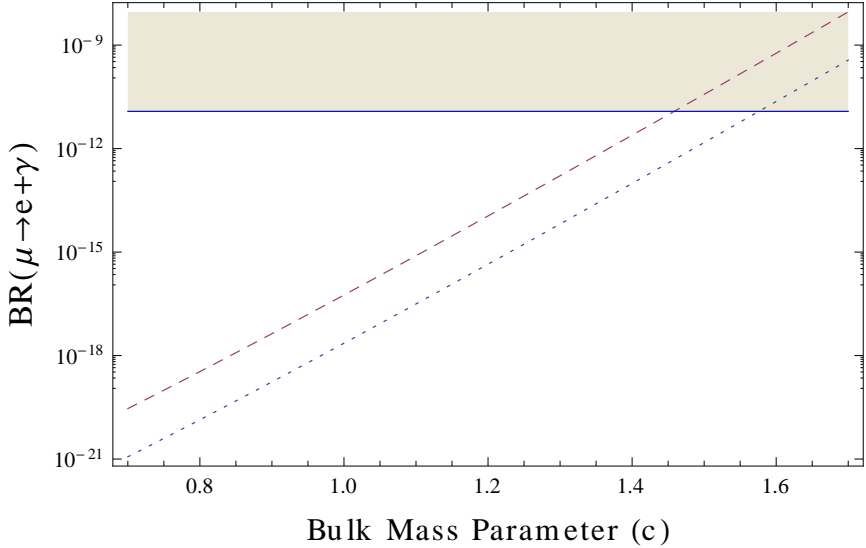}
\end{center}
\caption{The $\mu\rightarrow e+ \gamma$ Branching Ratio as a function
  of the Bulk
  Mass Parameter ($c$) for fixed values of the SM neutrino
  masses, $m_\nu=0.1$~eV. The dashed (dotted) line is for
  $\lambda_{\nl}=0.5\ (0.1)$ and the shaded region is experimentally excluded. We use
  $R^{-1}=1$~GeV,
  $k=1.5$~TeV and $k/M_*=1/2$ for the plot.}
\label{fig:mu_to_e_gamma_vs_c}
\end{figure}

We plot the $\mu\rightarrow e+\gamma$ branching ratio as a function of
the bulk mass parameter $c$ in
Figure~\ref{fig:mu_to_e_gamma_vs_c} for three
generations of bulk neutrinos with the sum again cut off at $n_*\simeq (M_* R)/\pi$. The SM neutrino masses are fixed
at $m_\nu=0.1$~eV and we use $k=1.5$~TeV, $R^{-1}=1$~GeV and
$k/M_*=1/2$ for the plot. The figure shows that $\mu\rightarrow
e+\gamma$ constraints are more stringent than the constraints previously considered;
values of $c\gtrsim 1.6\ (1.45)$ are already excluded for
$\lambda_{\nl}=0.1\ (0.5)$. Recall that $\lambda\in[10^{-2},0.1]$ when
$c\in[1.3,1.6]$ for
$\lambda_{\nl}=0.1$ and $m_\nu\simeq 0.1$~eV. Thus the interesting
region of parameter space in which the Yukawa coupling $\lambda$ is
$\mathcal{O}(0.1)$ corresponds to values of $c$ that are close to the
upper bound.

Other flavour changing processes like $\mu- e$ conversion in nuclei
and $\mu\rightarrow e+e^+e^-$ give similar constraints. The most
accurate experimental result in this regard comes from searches for
$\mu- e$ conversion in Ti. The branching ratio with respect to the
total nuclear muon capture rate is bound as $BR(\mu\
\mathrm{Ti}\rightarrow e\ \mathrm{Ti})<4.3\times 10^{-12}$~\cite{Dohmen:1993mp}. However,
this bound turns out to be less severe than the $\mu\rightarrow
e+\gamma$ constraint. For a single sterile neutrino, the resulting
constraint is $\left|\theta_{es}\theta^*_{\mu s}\  \sum_\ell
  |\theta_{\ell s}|^2\right|<1.3\times 10^{-3}\ (100\
\mathrm{GeV}/M_N)^2$, which is weaker due to the $\sim \theta_{\alpha
  s}^4$ dependence unless $M_N\gg 10$~TeV. We only consider
$M_n\lesssim1$~TeV in this work and thus find the $\mu\rightarrow
e+\gamma$ bound more useful.

In summary, we find that $\mu\rightarrow e+\gamma$ constraints on the
present model are the most severe and that these typically require
$c\lesssim1.6$ for $\lambda_{\nl}\simeq0.1$. Of interest is the fact
that larger values of $c$ correspond to larger values of the Yukawa
coupling $\lambda$ for a given value of $m_\nu$ (see
Figure~\ref{fig:yuk_vs_c_normal}). The more interesting regions of
parameter space thus correspond to values of $BR(\mu\rightarrow
e+\gamma)$ just below current limits. In particular, Yukawa couplings
of $\lambda\gtrsim 0.01$ correspond to $BR(\mu\rightarrow
e+\gamma)\gtrsim 10^{-14}$. The MEG collaboration is expected to
report new
results on the branching fraction for $\mu\rightarrow e+\gamma$ soon~\cite{MEG}. Given their targeted precision of measuring
$BR(\mu\rightarrow e+ \gamma)$ at the $\mathcal{O}(10^{-13})$ level,
MEG will have the potential to significantly improve the constraints
on the present model. In existing releases the MEG collaboration has
reported that the best value for the number of signal events in their
likelihood analysis is nonzero (specifically, three
events)~\cite{MEG}. Interpreted as a signal, this corresponds to
$BR(\mu\rightarrow e+ \gamma)=3\times 10^{-12}$, which is right in the
interesting range in the present
context (see Figure~\ref{fig:mu_to_e_gamma_vs_c}). We
eagerly await the release of more data by the MEG collaboration
to see if this
potential signal holds up. 
\section{Hidden Higgs Decays\label{sec:higgs_decay}}
The KK tower of sterile neutrinos has the potential to modify the
Higgs decay widths due to additional channels involving the sterile
states. The Higgs couples to the steriles via the UV localized
Yukawa interaction
\bea
S&\supset &-\frac{\lambda}{\sqrt{M_*}} \int d^5x\sqrt{-g_{uv}} \ \bar{L}
H N_{R}\ \delta(z-k^{-1})\ ,
\eea 
where $L$ is a lepton doublet and $H$ is the SM scalar doublet. In a
general 4D theory, the SM Higgs will decay into neutrino pairs $h\rightarrow
\nu N$ whenever the SM neutrinos Yukawa-couple to one or more light
sterile neutrinos. However, the demand $m_\nu\lesssim
\mathcal{O}(\mathrm{eV})$  typically
forces the Yukawa couplings to be tiny as one expects $\lambda_s\simeq
\sqrt{m_\nu M_N}/\langle H\rangle$ for seesaw neutrino masses;
$M_N\sim$~GeV gives $\lambda_s\simeq 10^{-7}$ for $m_\nu\simeq
0.1$~eV and the corresponding modification to the Higgs width is
therefore negligible. 

In the present context, the Yukawa couplings need not be tiny as
$m_\nu$ is suppressed by warping (or in the dual 4D theory,
fundamental/composite mixing) in addition to the usual seesaw suppression.  The action can be written as
\bea
S&\supset &-\frac{\lambda}{\sqrt{M_*}} \int d^5x\sqrt{-g_{uv}} \ \bar{L}
H N_{R}\ \delta(z-k^{-1})\nonumber\\
&= &-\sum_n  \int d^4x\ \frac{m_n^{\D}}{\langle H\rangle}\  h\bar{\nu}_L
\nu^{(n)}_{R}\  +\ \dots\nonumber\\
&= &- \int d^4x\ \
h\bar{\nu}_L\left\{\lambda_0^{\mathrm{eff}}N_{0R} +\sum_{n>0}\lambda_n^{\mathrm{eff}}\left(N_{+n L}^c +N_{-n L}^c\right)\right\}\ +\ \dots \nonumber
\eea 
where we define $\lambda_0^{\mathrm{eff}}\simeq\sqrt{m_\nu M_{00}}/\langle H\rangle$, use the fact that $\nu^{(n) c}_R\simeq(N_{+n }+N_{-n})_L/\sqrt{2}$, and  write the effective Yukawa coupling for the $n$th KK mode as
\bea
\lambda_n^{\mathrm{eff}}\equiv \frac{m_n^{\D}}{\sqrt{2}\langle H\rangle} \simeq 2\sqrt{|\lambda_{\nl}|}\times \frac{10^{-5}}{\langle H\rangle R}\left(\frac{m_\nu}{10^{-1}\ \mathrm{eV}}\right)^{1/2} \left(\frac{\mathrm{GeV}}{R^{-1}}\right)^{1/2}\left(\frac{M_n R}{2}\right)^{c}.
\eea
If one adds a sterile neutrino $N$ with Yukawa coupling $\lambda_s$ to the SM, the decay width for the Higgs to two neutrinos is:
\bea
\Gamma_N(h\rightarrow \nu\ N)&=& \frac{\lambda_s^2 m_h}{8\pi}\left(1-\frac{M_N^2}{m_h^2}\right)^2
\eea
where we neglect $m_\nu\ll M_N$. Neglecting the phase space factors, we can apply this result to the present model:
\bea
& &\Gamma(h\rightarrow \mathrm{invisible})\equiv 9\times\sum_n\Gamma(h\rightarrow \nu\ N_{n})\simeq \frac{9 m_h}{8\pi}\times    \left\{(\lambda_0^{\mathrm{eff}})^2+\sum_{n>0}2 (\lambda_n^{\mathrm{eff}})^2\right\},
\eea
where the sum  is cut off in the UV at $n_h\simeq m_hR/\pi$.

\begin{figure}[ttt]
\begin{center}
        \includegraphics[width = 0.7\textwidth]{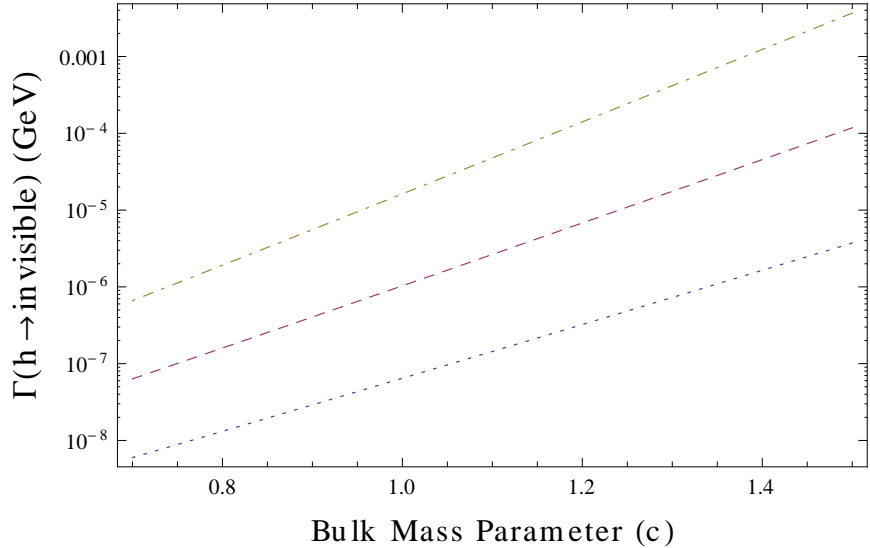}
\end{center}
\caption{The Invisible Higgs Decay Width as a function of the Bulk
  Mass Parameter ($c$) for fixed values of the SM neutrino
  masses, $m_\nu=0.1$~eV. The  dotted (dashed, dot-dashed) line is for
  $m_h=150\ (300,\ 600)$~GeV with $\lambda_{\nl}=0.5$ and $R^{-1}=1$~GeV.}
\label{fig:Higgs_width_vs_c}
\end{figure}

This invisible width is plotted as a function of the bulk mass
parameter $c$ in Figure~\ref{fig:Higgs_width_vs_c}. Observe that it is
significantly larger than that obtained in a standard low-scale seesaw
with order GeV sterile neutrinos (e.g., the $\nu$MSM), for which one
has $\Gamma_N\sim(\lambda_s^2/8\pi)m_h\sim10^{-12}$~GeV for $m_h\sim100$~GeV. Noting that
the width for SM Higgs decays is $\Gamma_{\mathrm{SM}}\sim
\mathcal{O}(10^{-3})\ [\mathcal{O}(100)]$~GeV for $m_h=150$~GeV
$[600$~GeV], the invisible branching fraction is
$\Gamma_{\mathrm{inv}}/\Gamma_{\mathrm{SM}}\lesssim 10^{-3}\
[10^{-4}]$ for the values shown in the figure. For a lighter Higgs
mass with larger values of $c$ the invisible
branching fraction is on the order of the two-photon width,
$BR_{\mathrm{inv}}\sim BR(h\rightarrow 2\gamma)$. Despite its small
width, the two-photon channel is known to be useful if the Higgs is
light, due to the relatively clean signal. However, invisible decays
are not as readily discerned from the data. The LHC can ``observe''
invisible Higgs decays, but only for much larger branching fractions
of roughly $13\%$ ($5\%$) for $10~fb^{-1}$ ($100~fb^{-1}$) of
data~\cite{Eboli:2000ze}. We conclude that, despite a significant
increase relative to other models with a low-scale seesaw, the width
for Higgs decays to neutrinos remains too small to be observed at the
LHC.  The ILC would have a better discovery (and limits) reach for
invisible Higgs decays, possibly achieving discovery (limits) for
invisible branching fractions of order $1\%$
($0.1\%$)~\cite{Richard:2007ru}.  

We have assumed ``natural values'' for the mixing angles in this
analysis; e.g., the natural value for the mixing angle between a
standard seesaw pair $\nu_L$ and $N_s$ is $\theta_s\simeq
m_\nu/M_N$. However, the
mixing between the SM neutrinos and the heavy neutrinos need not be as
small as this natural value if there are cancellations amongst the
parameters in the seesaw formula~\cite{de Gouvea:2007uz}. An enhanced
mixing angle means the Dirac Yukawa-coupling can be larger, leading to
an enhancement of the hidden Higgs decay width~\cite{de
  Gouvea:2007uz,Chen:2010wn}. Such cancellations could be considered in
the present context and the hidden Higgs width can be similarly
enhanced. 

We also note that the total invisible width can potentially be
higher than the neutrino width. Stabilization of the length of the
extra dimension via the Goldberger-Wise mechanism, for example, could open
additional invisible channels. The Higgs can couple to, and possibly
mix with, the Goldberger-Wise scalar and thus the overall invisible width could
increase. This is beyond the scope of the present work but we
hope to consider it in the future.

\section{Conclusion\label{sec:conc}}
The mini-seesaw mechanism achieves light SM 
neutrino masses by combining naturally suppressed
Dirac and (sterile) Majorana mass-scales together in a low-scale seesaw
mechanism. It employs a truncated (``little'') warped space 
that, via AdS/CFT, is dual to a 4D theory possessing conformal symmetry in some window, $M_*> E>R^{-1}$,
with $M_*\ll M_{Pl}$.  The model generates light SM neutrino masses without recourse to a large UV scale, and thus offers hope that the mechanism of neutrino mass generation may be explored in feasible experiments. 

A key feature of this approach is the existence of a  tower of light
sterile-neutrinos that mix with the SM. We have considered the
detailed bounds on these light neutrinos from processes like
$\nubb$-decay, $\mu\rightarrow e+\gamma$ and invisible $Z$-decays. We
find that the most stringent constraints come from FCNC processes like
$\mu\rightarrow e+\gamma$. Nonetheless we have shown that viable
parameter space exists in which the input Yukawa couplings can be
sizable. The model therefore provides a viable explanation for the
lightness of the known SM neutrinos. Furthermore, the $\mu\rightarrow
e+\gamma$ branching fraction lies just below current experimental
bounds for interesting regions of parameter space and, importantly, is
within the region to be probed by the forthcoming MEG experiment. We
look forward to the MEG results in the hope that (indirect) evidence for the origin of neutrino mass is revealed.
\section*{Acknowledgements\label{sec:ackn}}

The authors thank W.~Rodejohann and A.~Watanabe for helpful
discussions. MD was supported by the IMPRS-PTFS. DG was supported by the Netherlands Foundation 
for Fundamental Research of Matter (FOM) and the Netherlands Organisation 
for Scientific Research (NWO).
\appendix
\section{Appendix: Mass Mixing\label{sec:mass_mixing}}
\appendix
To $\mathcal{O}(M_{nn'}/M_n)$ the matrix that gives the heavy eigenstates is: 
\bea
U_{KK}=\left(\begin{array}{cccccc}
1& -\frac{1}{\sqrt{2}}(M_{01}/M_{1})&\frac{1}{\sqrt{2}}(M_{01}/M_{1})& -\frac{1}{\sqrt{2}}(M_{02}/M_{2})&\frac{1}{\sqrt{2}}(M_{02}/M_{2}) &\cdots\\
0&\frac{1}{\sqrt{2}}(1-\frac{M_{11}}{4M_1})&\frac{1}{\sqrt{2}}(1+\frac{M_{11}}{4M_1})& -\frac{M_{12}}{\sqrt{2}}\frac{M_2}{M_2^2-M_1^2}&\frac{M_{12}}{\sqrt{2}}\frac{M_2}{M_2^2-M_1^2}&\cdots\\ 
M_{01}/M_{1}&\frac{1}{\sqrt{2}}(1+\frac{M_{11}}{4M_1})&\frac{1}{\sqrt{2}}(-1+\frac{M_{11}}{4M_1})&-\frac{M_{12}}{\sqrt{2}}\frac{M_1}{M_2^2-M_1^2}&-\frac{M_{12}}{\sqrt{2}}\frac{M_1}{M_2^2-M_1^2}&\cdots\\
0& -\frac{M_{12}}{\sqrt{2}}\frac{M_1}{M_1^2-M_2^2}&\frac{M_{12}}{\sqrt{2}}\frac{M_1}{M_1^2-M_2^2}&\frac{1}{\sqrt{2}}(1-\frac{M_{22}}{4M_2})&\frac{1}{\sqrt{2}}(1+\frac{M_{22}}{4M_2})&\cdots\\
M_{02}/M_{2}&-\frac{M_{12}}{\sqrt{2}}\frac{M_2}{M_1^2-M_2^2}&-\frac{M_{12}}{\sqrt{2}}\frac{M_2}{M_1^2-M_2^2}&\frac{1}{\sqrt{2}}(1+\frac{M_{22}}{4M_2})&\frac{1}{\sqrt{2}}(-1+\frac{M_{22}}{4M_2})&\cdots\\
\vdots&\vdots&\vdots&\vdots&\vdots&\ddots
\end{array}\right).\nonumber
\eea 




\end{document}